\documentclass[a4paper,fleqn,usenatbib]{mnras}

\usepackage[T1]{fontenc}
\usepackage{ae,aecompl}

\usepackage{graphicx}	% Including figure files
\usepackage{amsmath}	% Advanced maths commands
\usepackage{amssymb}	% Extra maths symbols

\title[Model for relativistic shocks]{Particle acceleration, magnetization and radiation in relativistic shocks}

\author[Evgeny Derishev and Tsvi Piran]{
Evgeny V. Derishev$^{1}$
and Tsvi Piran$^{2}$
\\
$^{1}$Institute of Applied Physics, Russian Academy of Science, 46 Ulyanov Street, 603950 Nizhny Novgorod, Russia\\
$^{2}$Racah Institute for Physics, The Hebrew University, Jerusalem, 91904, Israel
}

\date{Accepted XXX. Received YYY; in original form ZZZ}

\pubyear{2015}

\begin{document}
\label{firstpage}
\pagerange{\pageref{firstpage}--\pageref{lastpage}}
\maketitle

\begin{abstract}
The mechanisms of particle acceleration and radiation,
as well as magnetic field build up and decay in relativistic
collisionless shocks are  open questions with important
implications to various phenomena in high energy
astrophysics.
While the Weibel instability is possibly responsible
for magnetic field build up and diffusive shock
acceleration is a model for acceleration,  both have problems and current PIC simulations show that
particles are accelerated only under special conditions
and the magnetic field decays on a very short length
scale. We present here a novel model for the  structure and the emission of highly
relativistic
collisionless  shocks. The model takes into account (and is based on)
non-local energy and momentum transport across the shock front via
emission and absorption of high-energy photons. This leads to a
pre-acceleration of the fluid and pre-amplification of the magnetic
fields in the upstream region. Both have drastic implications on the
shock structure. The model
explains the persistence of the shock generated magnetic field at
large distances from the shock front.  The dissipation of this magnetic field
results in a continuous particle acceleration within the downstream
region.
A unique feature of the model is the existence of an ``attractor'', 
 toward which any shock will evolve.
The model is applicable  to  any relativistic shock, but its
distinctive features show up only for sufficiently large compactness.
We demonstrate that prompt and afterglow Gamma-Ray Bursts'  shocks
satisfy the relevant conditions and we compare  their observations
with the predictions of the model. 
\end{abstract}


%%%%%%%%%%%%%%%%%%%%%%%%%%%%%%%%%%%%%%%%%%%%%%%%%%

%%%%%%%%%%%%%%%%% BODY OF PAPER %%%%%%%%%%%%%%%%%%

\section{Introduction}

Relativistic outflows are ubiquitous in extreme astrophysical phenomena. Specifically, they arise in  Gamma-Ray Bursts (GRBs), Active Galactic Nuclei (AGNs) and  microquasars. Emission from such outflows 
originates at large distances from the central engine driving the outflow, most likely due to relativistic collisionless shocks that take place either within the outflows (internal shocks) or at the interface with surrounding medium (external shocks).    
According to the standard picture the shocks amplify magnetic fields and accelerate particles and these emit the observed radiation. The magnetic field energy density and the electrons' energy density are characterized by equipartition parameters (typically assumed to be not much smaller than unity) that relate these energy to the total energy in the downstream. 

While synchrotron emission characterized by such parameters works rather well in some cases (e.g. GRB afterglow) both observational and theoretical problems arise. 
Within GRB prompt emission this model is inconsistent 
with the hard low energy spectrum observed at times \cite{Cohen+97,Preece+98,Ghisellini+00,Gruber+14}. The apparent lack of high-energy inverse Compton (IC) component \cite{LATGRB13} is also a puzzle. 

On the theoretical side it was expected that 
the magnetic field at the front is comparable to the equipartition value: even if there is no field in the upstream, it will be generated through the Weibel instability. This was predicted theoretically \cite{TheorWeibel1,TheorWeibel2} and later confirmed by numerical simulations 
\cite{NumericWeibel}. However, these  Particle-in-cell (PIC) simulations indicate (in accord with early semi-analytic considerations \cite{Gruzinov}), that the generated magnetic field decays rather quickly on the scale of the downstream skin depth. It is a matter of debate, how strong could be the magnetic field in the downstream far away from the shock front. Lacking a strong magnetic field downstream it is unclear how can this region radiate. 
Additionally, it is expected that particles are accelerated in collisionless shocks via the Diffusive Shock Acceleration (DSA) mechanism \cite{DSA}. However the same PIC simulations show that while DSA operates well  if there is no magnetic field in the upstream or if the magnetic field is parallel to the shock normal, it becomes ineffective  in the presence of even a modest perpendicular magnetic field (see \cite{PICsimulReview} and references therein).

These recent  theoretical developments  question the ability of relativistic collisionless shocks to generate the observed emission and have led to numerous suggestions of alternative 
scenarios. For example,  IC-modified  thermal emission from a fireball 
\cite{RM05,Ryde05,Peer06,Photosph1,Photosph2}, magnetic reconnection within a Poynting flux dominated outflow \cite{LyutikovBlandford03} and hadronic  models (e.g. \cite{Murase+12}) have been proposed within the context of GRBs' prompt emission.  However these models have their own share of both phenomenological and theoretical problems. We present here a novel model of relativistic shocks that consistently treats both particle acceleration and radiation and resolves the above mentioned theoretical problems. The model is applicable to shocks with moderate magnetization, i.e., excluding Poynting-flux dominated outflows.

Other attempts to resolve these problems that have taken into account the influence of the shock on the upstream include: heating, pre-acceleration, build-up and distortion of the magnetic field, and possibly changes in the composition. Generally speaking 
there are two classes of such shock modification models. The first includes  energy and momentum transport by fast particles or waves that propagate ahead of the shock front (see, e.g., \cite{ShockModif1,ShockModif2,ShockModif3,MiloNakar,MedvZak}).  Another option, is  electron-positron pair production in the upstream. This latter possibility has been suggested to arise if enough  photons are backscattered by the surrounding matter. These backscattered photons annihilate some of the similar-energy outgoing photons  (see, e.g., \cite{pairload,PairLoad2}) creating pairs.  For sufficiently large external scattering this leads to a runaway increase of pair density in the upstream  modifying its structure. 
This is a non-local mechanism and,  if it works, it always  becomes the dominant one. However it depends critically on  the scattering opacity of the external medium and this  limits its applicability  to external shock in some GRBs.

We propose here a different modification of the shock model. We suggest that ``intrinsic" annihilation of outgoing high-energy and low-energy photons  will also  create pairs in the upstream and modifies this region.  
The key difference between this and the ``standard" pair-loading scenario  is that the pair-creation is ``intrinsic", i.e.  it 
does not involve scattering by an external medium.  As such the process is generic and will take place in any relativistic shock. However, now a source of high-energy photons in the downstream is needed. These 
high-energy photons arise due to  inverse Compton scattering. 
In this case the length scale for the  feedback between the downstream and upstream regions is set by the parameters of the shock. This allows for self-tuning of the shock structure (including self-consistent build up and decay of the magnetic field, as discussed later). As this is a long range process the modification of the upstream takes place on a much longer scale than the modification that arises from fast particles that are moving ahead of the shock front creating an important long range structure in the magnetic field.  
 
A second key idea in our model is the modification of the magnetic field by the pairs. The pairs that are produced in the upstream form an anisotropic lepton distribution which, in turn, leads to a pre-amplification of the magnetic field.   
This pre-amplification  results in long wavelength modes. The magnetic field, which is further amplified at the shock front itself, decays in the downstream on a scale much larger than the skin depth in the downstream region. 
This picture is in conflict with PIC simulations that show a rapid decay of the magnetic field  (e.g., \cite{PICsimulReview}). However current PIC simulations don't include the long-range momentum exchange processes discussed here. This process  is 
mediated by high energy photons that are produced in the downstream and absorbed creating pairs in the upstream.
Hence these PIC simulations cannot capture this feature of our model.

The energy released by the dissipating magnetic field is inevitably transferred to charged particles and then to synchrotron and inverse Compton radiation. The
local distribution function of the leptons (and hence the local spectrum) is controlled by the magnetic field dissipation rate and it evolves together with the magnetic field as the shocked plasma advected downstream. Compared to the standard one-zone synchrotron shock model \cite{SSM1,SSM2}, this allows for additional flexibility in the emerging spectra \cite{MultiZone}. This also leads to a similar length scale for the magnetic field decay,  particle acceleration, and  the radiative cooling. Thus this model  does not suffer from  incommensurability of the magnetic field decay and radiative cooling scales (in our model they appear to be approximately equal). 

In order to avoid confusion let us explicitly list the key differences of our model from the pair loading scenario  \cite{pairload,PairLoad2}, mentioned above.  To distinguish between the two we denote this model as ``external'' pair loading scenario while the pair loading in our model is ``intrinsic''.  In the external pair loading scenario the target  photons are  outgoing {($\sim$ MeV)} photons that have been reflected by the surrounding matter.  These reflected photons annihilate outgoing {(also $\sim$ MeV)} photons. 
The  resulting pairs have a small Lorentz factor relative to the lab frame. 
In our intrinsic model, pair production takes place between high-energy (IC) photons and low-energy synchrotron ones that are both moving forward 
in the lab frame . The pairs are highly relativistic and beamed in the lab frame. The number of pairs created  does not depend on reflection from the external medium. Instead  it  is self-tuned to fit the shock's energy release.

{An extention of} the external pair loading scenario {suggests that the pairs may also influence the magnetic field in the upstream (\cite{Ramirez-Ruiz+}). However}, the delay between the onset of pair loading and arrival of the shock front is set by the shock propagation timescale. This scale is independent and much larger than the intrinsic shock timescales, {especially} the magnetic damping timescale.   Thus,  any  magnetic turbulence generated in the upstream by these pairs is damped and   will have a very limited effect on the magnetic field at the shock front. In contrast, in our intrinsic model the upstream modification {scale} is controlled by {the upstream absorption timescale, which in our model is proportional to} the {downstream} cooling timescale {(as explained in the paper),} 
and it controls the magnetic damping timescale  so all {scales} vary in the same way.

The only way to fully explore our model is through numerical PIC simulations that include this long-range energy and momentum exchange  between the downstream and the upstream. However, a simplified phenomenological version of the model can yield important analytic insights towards general features of relativistic shocks and their broad-band spectra.
We describe and analyze such  a model here. In order to advance with an analytic solution, we introduce several assumptions. The assumptions are not required by the model in a general sense; it works just as well when they are not satisfied. But they greatly reduce the model's complexity and yet are reasonable for the shocks expected in GRBs and in AGNs. In particular  we analyze a simplified, phenomenological model, in which we assume the magnetic field decay law rather than calculate it self-consistently. This keeps the problem tractable and enables us to obtain analytic results regarding the structure of relativistic shocks and their emission.

When applying these ideas to GRBs, it should be noted that the mechanisms behind the prompt and the afterglow emission may be different. 
In the following we don't attempt to provide a complete phenomenological solution to either prompt or afterglow emission. In fact the current status of the development of the model, as exposed here, is probably not suitable for this task as yet. Instead, we point out in numerous places in the text, the consistency of our estimates with typical parameters, believed to be applicable to both internal and external shocks in GRBs.  While the full observational implications of the model are not clear yet we explore various pros and cons of this model when confronted with GRB observations in the concluding paragraph.

We begin, in \ref{sec:frame}, with  a step-by-step description of model's physics. 
In \ref{sec:scales} we discuss the spatial scales relevant to the problem and their hierarchy. 
Each of the model's building blocks is then substantiated and discussed in greater detail in one of the subsequent sections. We conclude in \ref{sec:discussion} summarizing the basic features of this model and some of its observational implications.

\section{Model's framework}
\label{sec:frame}

\begin{figure}
\includegraphics[width=\columnwidth]{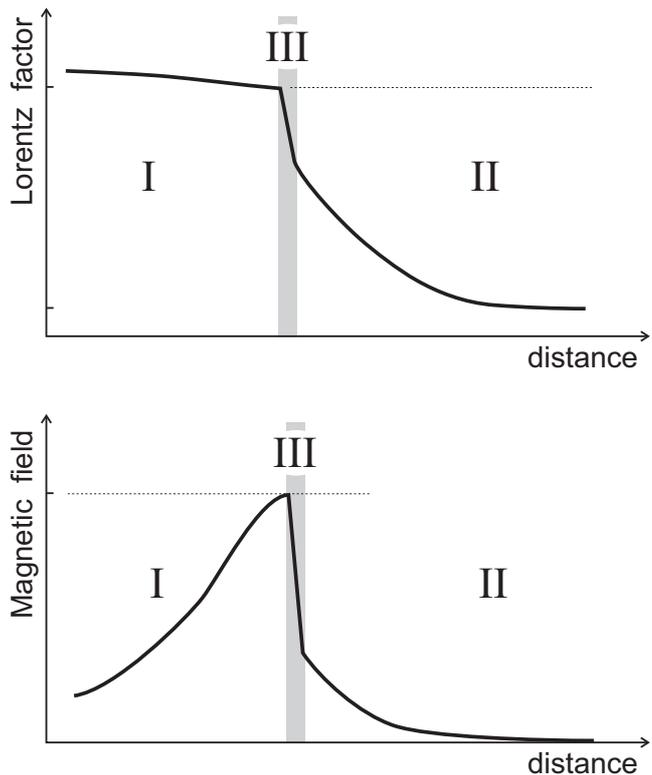}
\caption{A schematic description of the shock structure
with three zones (downstream -  I , pre-accelerated upstream - II and the shock front - III). 
Shown are the bulk Lorentz factor
in the laboratory frame (upper panel) and the comoving-frame
magnetic field strength (lower panel) as functions of distance.
The distance is measured along the shock normal and counter to the
shock velocity. The shock moves to the right.}
\label{modelscheme}
\end{figure}

{\it Shock structure:} 
We begin with an outline of the physical framework of the model. 
We consider an ultra relativistic shock propagating into a surrounding cold region.
The shock is   divided into three regions, denoted in Fig.~\ref{modelscheme}  by Latin numbers. The downstream 
region (I), which we consider as the primary source of emission, is filled with hot
compressed plasma, where both electrons (with a possible admixture
of positrons) and protons (or ions more generally) are highly relativistic. 
The magnetic field gradually decays in this region
and it is this decay that heats the particles there. Energized electrons then emit both synchrotron and high-energy inverse Compton radiation.
Absorbed high energy photons from the downstream  lead to pre-acceleration at the upstream region (II) and to pre-amplification of the magnetic field there. 
The magnetic field is further amplified at the shock front (III) and particles are boosted once they cross it. 
We do not resolve the front  and we consider it as an infinitely thin
surface. Unlike the standard shock model in which both particle acceleration and magnetic field amplification take place  in this front,  its role is reduced here to the amplification of the magnetic field.   

We assume that the total thickness of the structure, shown in
Fig.~\ref{modelscheme}, is much less than the shock curvature
radius, so that the system is essentially one-dimensional. This curvature
radius, however, enters as a parameter that determines
the timescale for photon escape and hence the radiation density in
the emitting regions.

{\it Emission and absorption of high-energy photons:} 
The resulting radiation field is produced within the different components of the shock
either via synchrotron radiation or via IC. Both originate from the populations of relativistic electrons and positrons (whose  number and the distribution function differ from one point to another). The luminosities of the synchrotron and IC emission are approximately equal when the shock is in the fast cooling regime or close to it. This claim is substantiated in Sect.~\ref{sec:IC}.  

The estimated energy of the radiating electrons is often close to the value, where the Klein-Nishina effect becomes important. In Sect.~\ref{sec:tuning} we show that this is not a coincidence, but rather a natural result of shock parameters evolution. Under such conditions, absorption of the IC radiation inside the shock  
is important
(see Sect.~\ref{sec:IC} for quantitative estimates). Indeed, by definition at the Klein-Nishina cutoff the energy of the upscattered photons is just at the optimum for two-photon pair production with the low energy photons. As the cross-section for pair production is of the same order of the scattering cross-section. So, if cooling through IC emission is efficient, then two-photon absorption of the produced radiation is also efficient. 

{\it Upstream pre-acceleration: }
Some of the high-energy IC photons produced in the downstream region  interact  within the shock  region
 with low-energy synchrotron photons and produce pairs. This strongly influences  the shock's structure and shapes its evolution.  This process  links upstream to the downstream, so that the upstream is no longer  casually disconnected from the downstream. 
This non-local energy and momentum exchange between the downstream and the upstream is the backbone of our model. 

The secondary pairs transfer energy and momentum to the upstream, so that it starts accelerating long before the shock front arrives. As a result, the velocity jump across the shock front decreases, possibly to the extent that it becomes sub-relativistic, or even disappears if the pair loading is strong enough. At the same time, the downstream velocity relative to the shock front increases. The effect of upstream pre-acceleration is quantitatively analyzed in Sect.~\ref{sec:preacceleration}.

{\it Energization at shock crossing:  }
The decrease in the shock front speed and the increase in the downstream velocity relative to the shock front act in accord to make diffusive shock acceleration less efficient. 
Hence, unlike the conventional shock acceleration model (see \cite{SSMreview} for a review), we don't  postulate acceleration of particles at the shock front. Instead, two other mechanisms maintain the population of energetic emitting particles.

Once 
the energetic pairs, which have been produced in the upstream  cross the shock front, they are boosted by the Lorentz factor between the upstream and the shock front.  Thus, the energy of such pairs exceeds the energy of the original photons that have produced them.
This population of energetic electrons is the source of the next generation of photons that, in turn, will be again absorbed in the upstream producing even more energetic pairs. With each subsequent cycle, both the total and the individual energy of the particles involved into this process are multiplied by a large factor. This constitutes the converter acceleration mechanism 
\cite{ConvAcc,ConvAcc2} that  operates in every relativistic shock with a non-vanishing opacity for photon-photon interaction. Without special measures to counteract it, the converter acceleration is over-efficient and eventually it  leads to depositing all the available energy into the highest energy pairs, resulting in a cascade spectrum that is very different from the observed one. 
In Sect.~\ref{sec:UpstrAccel} we describe a self-regulating mechanism that keeps this process under control and limits it. The net result of the converter acceleration in our model is the formation of high-energy tail in the electron-positron distribution function. These high-energy electrons and positrons are injected into the downstream at the shock front and contain a sizeable, but not dominant, fraction of energy in radiating particles.

{\it Magnetic field generation and decay:}
In addition, the continuous injection of pairs maintains anisotropic particle distribution in the upstream, which gradually builds up  the magnetic field over a distance that is orders of magnitude larger than the plasma skin depth. 
The magnetic energy is therefore channelled to large-scale modes, which grow slowly, but also survive for long time in the downstream. The spatial scale for the magnetic field growth and decay is set by the absorption length of the IC radiation and it is comparable to the shock's dynamical scale (the timescale of shock's expansion, measured in the comoving frame). The estimates of the magnetic field decay time, as well as its correlation length and magnitude are presented in Sect.~\ref{sec:Bevolution}.

The long range decay of the magnetic field is a key feature in our model . As was noted earlier, the assumption of slow magnetic field dissipation apparently contradicts results of PIC simulations. However, this contradiction does not prove that either our model is wrong or that the simulations are incorrect. Instead, the simulations  deal with a different physical problem in which some of the crucial ingredients of our model are missing. In particular, the PIC simulations don't involve the radiation field and hence they  miss the gradual pre-acceleration of the fluid and the gradual build up of the magnetic field in the upstream region. These processes arise from absorption of photons produced by the downstream in the upstream region. Put differently, the PIC simulations deal with an almost  instantaneous build-up of the anisotropy in the particle distribution function that is followed by a rapid build up and a rapid decay of the magnetic field, whereas in our model the anisotropy is maintained by injection over a long range.

The longest available PIC simulations show that the magnetic field dissipation length increases (together with the field correlation length) as the particles gain energy at the shock and increase their mean free path \cite{ScaleEvolution}. 
The lesson to be learned from the simulations is that the magnetic field decays over a length, which is roughly equal to its  build-up length.
In our model, due to the energy and momentum exchange via the photons, the build-up length is many orders of magnitude larger than the plasma skin depth. We expect the decay to follow suit. 

{\it Particle acceleration due to the magnetic field decay:}
The energy
released from magnetic dissipation in the downstream heats the  plasma  and it  is channeled to accelerated particles. This results in  a distributed particle acceleration (heating) all over the downstream emitting region.

Once accelerated  the electrons and positrons lose energy to synchrotron and inverse Compton radiation. 
We assume that radiative cooling is fast, so that the electrons and positrons
emit locally the energy they receive from the dissipating
magnetic field. The average electron's Lorentz factor 
is determined, in this case, by the local balance between energy gain and losses. 
The assumption of fast cooling is not a necessary ingredient of the model and it is done only 
in order to simplify the estimates carried out here. 

We assume that the net effect of the  heating and the radiative losses is to
produce either a relativistic thermal distribution or a power law distribution with smoothed low-energy cut-off. 
These options are meant to deal with two opposite extremes: the first is for the case where heating occurs through a rapid succession of small energy gains, while the second corresponds to the case where heating results from rare leaps in energy, each one large compared to the average energy. 

Electrons heated by magnetic dissipation in the downstream produce the bulk of the shock's emission. In Sect.~\ref{sec:heating} we analyze the evolution of electron distribution function in the downstream. This evolution, as we will show, determines the spectral shape at frequencies below the synchrotron peak, whereas the peak itself is maintained by the process of pair production at the location, which corresponds to Comptonization at the verge of Klein-Nishina cutoff.

{\it Pair loading in the downstream and self-tuning:}
If the shock starts with such parameters, that the IC radiation is produced in the Klein-Nishina regime, then the secondary pairs can easily outnumber the primary electrons. Since the heating power is fixed by the magnetic field decay, a larger number of radiating particles implies that an average particle becomes less energetic and produces less energetic  IC photons, which are absorbed less efficiently. This feedback rapidly drives the shock's parameters to the point, where an average IC photon is just below the pair production threshold when interacting with the most abundant synchrotron photons.  The opacity for IC radiation drops and so does the pair production rate. Thus, the pair production rate stabilizes at the level where it exactly resupplies the particle loss into the downstream. The resulting steady state represents an attractor solution that exists for a wide range of parameters.  Interestingly, with reasonable parameters,  this attractor solution yields peak photon energies that are in the range of observed peak energies for GRBs'  emission. The self-tuning of relativistic shocks is discussed in Sect.~\ref{sec:tuning}.

{\it The different radiation components:}
Overall there are three distinct emitting regions. The primary (most powerful) source of the synchrotron and IC radiation is the downstream region of decaying magnetic field. In Sect.~\ref{sec:downstream_region} we calculate the spectrum of synchrotron emission from this region, but we do not discuss the spectrum of IC component in any detail. 
The most energetic pairs that were born upstream may
have enough time to cool radiatively before reaching the shock
front. These are responsible for formation of yet another, upstream, emission component. 

In general, the spectra and temporal evolution of downstream and upstream emission are very different because of different bulk Lorentz factors, which, in addition, is not uniform across the upstream.  Typically we expect the upstream emission component to be weak compared to the downstream one. However, this may change under certain conditions. As discussed in Sect.~\ref{sec:IC}, the upstream intercepts a fair fraction of radiation released by the downstream, and potentially a part of this may be re-radiated by the upstream itself.
Finally a third component is due to 
energetic leptons from the upstream that have been boosted while crossing the shock front and reached  the  downstream. There these pairs find themselves in strong magnetic field and quickly lose their energy, mostly through the synchrotron emission. This gives rise to a radiation component, which we call transitional since it is due to particles that transit from the upstream to the downstream and then radiate in the vicinity of the shock front. The contribution of the transitional emitting region to the synchrotron SED is a power-law tail at high energies, which is calculated in Sect.~\ref{sec:other_regions}. 
Again, we don't calculate the spectrum of IC component in the transitional radiation, but note that it is of relatively smaller importance because of the Klein-Nishina cutoff.  

\section{Length scales}
\label{sec:scales}

At a relativistic shock, where two plasma flows meet
and penetrate into each other, the magnetic field is generated via
a Weibel-type instability. This process proceeds rather quickly, 
{up to} the length of the order of relativistic
plasma skin depth (in the comoving frame)
\begin{equation}
\label{skin} l_{s} = \left( \frac{\Gamma m_{p} c^2}{4 \pi e^2
n_{p}} \right)^{1/2} = \frac{1}{n_{p}} \left( \frac{e_{t}}{4 \pi e^2}
\right)^{1/2},
\end{equation}
where $\Gamma$ is the shock Lorentz factor, $m_{p}$ the proton
(ion) mass, $e$ the elementary charge, $n_{p}$ the number density
of protons, and $e_{t}=\Gamma m_{p} c^2 n_{p}$ the total energy density, both in the shock-comoving
frame. 
For GRB external shocks, for example, the comoving length given by Eq.~\ref{skin} is
\begin{equation}
	l_{s} \simeq \frac{4\times 10^5\, {\rm cm}}{n_3^{1/2}} \ ,
\end{equation}
where $n$ is the number density of ambient medium in the lab frame and the subscript $_x$ denotes a quantity in cgs units of $10^x$. The density 
$n_3 \sim 1$ is  typical for a wind from a Wolf-Rayet star at the distance of $10^{16}$~cm.
The large-scale magnetic turbulence generated in the upstream slows down the Weibel instability and broadens the shock front, as discussed in Sect.~\ref{sec:Bevolution}, but its width remains many orders of magnitude smaller than any other spacial scale in the system.

Numerical PIC simulations show that the magnetic field initially takes
up a substantial fraction of the shocked plasma energy density. It has  
a spacial scale of the order of $l_{s}$. As the plasma recedes
from the shock front, the field's spacial scale increases and its
strength decreases (see \cite{PICsimulReview} and references therein). 
This results from rearrangement of the
magnetic field lines and possibly from reconnection that 
generates strong local electric fields. Charged particle are
accelerated when they randomly encounter these electric fields. In this way  energy is transferred from the decaying magnetic field to the electrons and positrons. All relativistic particles of the same charge are energized on equal grounds, so that the non-radiating
protons would receive (and waste) half of the magnetic decay
energy, unless electrons and positrons greatly outnumber protons and take over energy expenditures.
As we show later, our model fits the latter case. Since
there is no reason for the electron distribution to differ from
positron distribution, we will make no difference between
electrons and positrons, and the name ``electrons" hereafter
denotes both.

We will characterize the length scale for the magnetic field dissipation, $l_d$, as
\begin{equation}
\label{Bdecay} l_{d} = \lambda l_{s},
\end{equation}
where $\lambda \gg 1$ is the model parameter. Though the actual
value of this parameter is unknown, it is likely very large, 
exceeding unity by many orders of magnitude. 
The decay length of the magnetic field is set up by the spatial scale of the currents rather than by the plasma skin depth. The low-$k$ modes of the magnetic field generated in the upstream  increase the correlation length of the downstream magnetic field and hence the decay length. 
Thus, the magnetic field decay length is determined by the length scale on which the IC photons are absorbed in the upstream, 
\begin{equation}
l_{abs} = R/\Gamma/\tau_{c},
\end{equation} 
where $\tau_{c}\equiv \sigma_{\gamma\gamma} n_{ph} {R}/{\Gamma}$ is the optical depth, $n_{ph}$ is the target photon number density and $\sigma_{\gamma\gamma}$ is the effective two-photon pair production cross section. To evaluate the absorption scale we will need to wait to  section \ref{sec:downstream_region}. 
Typically $l_{c} < l_{abs}  \approx l_{d}$.

A third important scale  is the electron cooling
length. Moving into the downstream with a bulk velocity $\beta_{d} c$ (for an unmodified 
relativistic shock $\beta_{d} = 1/3$), the electrons cool over the distance
\begin{equation}
\label{CoolingScale} l_{c} = 
%\frac{\epsilon \beta_{d} c}{\dot{E}} =
\frac{3 \beta_{d} m_{e} c^2}{4 \sigma_{T} \gamma (1+y) e_{B}},
\end{equation}
where $m_{e}$ is the electron's rest mass and $\gamma$ its Lorentz factor, 
$\sigma_{T}$ is the Thomson cross-section, $e_{B}$ the
magnetic field energy density and  $y$ is the Compton parameter. 
It is convenient to define the average Compton parameter, $\bar{y}$:
\begin{equation}
\label{avgCompton} (1+\bar{y}) \int_1^{\infty}\, \gamma^2 f_{e}\,
{\rm d}\gamma \equiv \int_1^{\infty}\, \gamma^2 (1+y) f_{e}\, {\rm
d}\gamma\, ,
\end{equation}
where $f_{e} (\gamma)$ is the electron distribution function. The
actual value of $\bar{y}$ turns out to be
insensitive to the model details; normally $\bar{y}$ is
about a few.
Numerically (for GRBs' external shocks), the cooling distance is 
\begin{equation}
l_{c} \simeq 
\frac{5\times 10^{9}}{\gamma_3 \Gamma_2^2 n_3 \epsilon_{B}}\,
\mbox{cm},
\end{equation}
where $n$ is the number density of ambient medium in the lab frame and the magnetic equipartition parameter is defined as $\epsilon_{B} \equiv e_{B}/e_{t}$.

It is instructive to compare
the cooling distance with the plasma skin depth:
\begin{equation}
\frac{l_{s}}{l_{c}} \simeq \frac{8}{9 \beta_{d}}  
\gamma (1+y) \epsilon_{B} \left( 4\pi
r_{e}^3 n_{p} \right)^{1/2} \Gamma^{3/2} 
\left( \frac{m_{p}}{m_{e} } \right)^{3/2},
\end{equation}
where $r_{e}=e^2/(m_{e}c^2) \simeq 2.8 \times 10^{-13}$~cm
is the classical electron radius. 
The {shock-front width (roughly equal to the skin depth)} is always orders of magnitude smaller than the cooling
distance. For example, in GRBs' external shocks
\begin{equation}
	\frac{l_{s}}{l_{c}} \sim 10^{-4}\, \gamma_3 \Gamma_2^2 n_3^{1/2} \epsilon_{B}.
\end{equation}
A mildly relativistic shock would require densities
typical for condensed matter to make these scales comparable. {It makes possible modeling of radiatively efficient shocks in laboratory laser plasma.}

We assume  the following hierarchy of spacial scales:
\begin{equation}
\label{hierarchy} l_s \ll l_{c} \ll l_{d} \leq R/\Gamma,
\end{equation}
where $R$ is the shock radius and $R/\Gamma$ is the comoving shock
scale. The right inequality ensures that the magnetic field
transfers substantial part of its initial energy to particles,
while the central one guarantees that the fast cooling
approximation is adequate. The right inequality inevitably breaks down at some point in the course of the deceleration of an external shock, and the second one is more like $l_{c} \lesssim l_{d}$ when it comes to a self-consistent solution (we will elaborate on this in Sections \ref{sec:tuning} and \ref{sec:downstream_region}).

\section{Distribution function of electrons in the downstream}
\label{sec:heating}

When the decaying downstream magnetic field heats electrons and positrons, which are in the fast cooling regime, their average energy is
set by the local balance between the heating rate per unit volume and
the average emissivity of the electrons. The former equals the decrease
of magnetic field energy density in the case where electrons
greatly outnumber protons and half of that otherwise\footnote{Highly relativistic electrons and protons, having equal charge, should gain the same energy when accelerated in local electric fields, which result from the magnetic field dissipation.}, and the latter is determined by the
shape of electron distribution over their energies. This shape
depends on the amplitude distribution of kicks that the  electrons
receive when passing through the electric fields. As we do not explore the details of the dissipation process of the magnetic field, the corresponding statistic of
kicks is unknown and some a priori assumption about the
electron/positron distribution is necessary. Simple and
reasonable choices are  either relativistic thermal distribution
\begin{equation}
\label{distribution} f_{e} (\gamma) = \frac{27 }{2\gamma_0^3}\,
\gamma^2 \exp (-3\gamma/\gamma_0)\, n_{e} \ , 
\end{equation}
or a power law distribution
\footnote{While smooth at $\gamma_0$, this distribution is effectively similar to the truncated power-law with a minimal Lorentz factor,  $\gamma_{min}$, commonly used in 
GRB modeling. }
\begin{equation} \label{distribution_PL} 
\begin{split}
f_{e} (\gamma) = \frac{p(p^2 - 1)}{2}\, (C \gamma_0)^{p - 1} 
\frac{\gamma^2}{(C \gamma_0 + \gamma)^{p+2}}\, n_{e}; \\
C = \frac{p - 2}{3}, \quad p>3.
\end{split}
\end{equation}
For both distributions the mean electron energy is $\gamma_0 m_e c^2$. The bulk of synchrotron emission is due to
electrons with $\gamma \sim (5/3) \gamma_0$ for the thermal distribution and 
$\gamma \sim (5(p-2)/3(p-3)) \gamma_0$ for the power-law one.

The inequality (\ref{hierarchy}) implies that the energy change due to
adiabatic expansion or compression is negligible. Then, the
equilibrium average Lorentz factor $\gamma_0$ varies with the
distance from the shock front following the changes in the local
energy density of the magnetic field and can be calculated from
the balance between acceleration and losses:
\begin{equation}
\frac{4}{3} \sigma_{T} e_{B} \int_1^{\infty}\, \gamma^2 (1+y)
f_{e} (\gamma, \gamma_0) \, {\rm d}\gamma = -\beta_{d}\, \frac{\partial
e_{B}}{\partial r}\, ,
\end{equation}
where $r$ is the distance from the shock front.  
Substitutions from Eqs. (\ref{avgCompton}) and one of 
(\ref{distribution}), (\ref{distribution_PL}) result in
\begin{equation}
\label{balance} \frac{4}{3}\, \left<\gamma^2\right> \sigma_{T} n_{e} e_{B}
(1+\bar{y}) = -\beta_{d}\, \frac{\partial e_{B}}{\partial r}\, .
\end{equation}
The average of the Lorentz factor squared is 
$\left<\gamma^2\right> = (4/3) \gamma_0^2$ for the thermal distribution and  
$\left<\gamma^2\right> = (4(p-2)/3(p-3)) \gamma_0^2$ for the power-law distribution.
The function $\gamma_0(r)$, 
can be determined as soon as the magnetic field decay law
$\partial e_{B}/ \partial r$ is known. Most reasonably, $\gamma_0$ is a 
monotonically decreasing function
of the distance from the shock and hence its value near the shock front determines
the bulk properties of emission spectrum.

\section{Emission and absorption of Inverse Compton radiation}
\label{sec:IC}

Equation \ref{balance} allows us to find the synchrotron emission
coefficient
\begin{equation}
\label{SyEmissivity} j = -\frac{\beta_{d} c}{4 \pi (1+\bar{y})}
\frac{\partial e_{B}}{\partial r}
\end{equation}
and then the intensity of synchrotron radiation at the front of
plane-parallel shock is
\begin{equation}
\label{SyIntensity} I(\theta) \simeq \frac{1}{\cos(\theta)}
\int_0^{\infty} j\, {\rm d} r \simeq \frac{\beta_{d} c\, e_{B}(0)}{4 \pi
(1+\bar{y}_0) \cos(\theta)}\, .
\end{equation}
Here $\theta$ is the angle between shock normal and the line of
sight. The first equality in Eq.~(\ref{SyIntensity}) {would be exact in the downstream frame assuming that the downstream velocity is constant, but in the shock-front frame it} is approximate because the downstream is moving, although at a non-relativistic velocity, and the Doppler boosting makes beam pattern to appear anisotropic in the shock-front frame even if it is isotropic in the comoving frame. 

According to Eq. (\ref{SyIntensity}), the intensity diverges as
$\theta$ tends to $\pi/2$. This is an artifact of the plane geometry approximation. However, the  shock  has a finite curvature,
so that the maximum intensity is limited. This can be taken into account by introducing the geometrical factor $\Lambda \simeq \ln (R/l_{d}/\Gamma)$. The approximate
energy density of the synchrotron radiation at the shock front is
\begin{equation}
e_{sy} = \frac{2\pi \Lambda I(0)}{c} =
-\frac{\Lambda \beta_{d}}{2} \int_0^{\infty} \frac{\partial e_{B}/\partial
r}{1+\bar{y}}\, {\rm d} r \, .
\end{equation}

The energy density of radiation produced by a geometrically thin
shock depends logarithmically on the distance from the shock front
and hence can be considered constant. If, in addition, the
Klein-Nishina cutoff does not appear  near to a local SED
maximum (usually it is the synchrotron SED maximum), 
then the fraction  of the radiation density $e_{r}$ that accounts 
for the  inverse Compton losses is also nearly constant.
Substituting in the above integral $\bar{y}$ with $e_{r}/e_{B}$,
and assuming $e_{r}=const$, we obtain
\begin{equation}
\label{Wsy} \frac{e_{sy}(0)}{e_{B}(0)} = \frac{\Lambda \beta_{d}}{2}\,
\left[ 1 - \bar{y}_0 \ln \left( 1+ \frac{1}{\bar{y}_0} \right)
\right]\, ,
\end{equation}
where the argument (0) stands for the shock front ($r=0$) 
and $\bar{y}_0=\bar{y}(0)\simeq e_{r}(0)/e_{B}(0)$. 

Equation (\ref{Wsy}) can be solved for a number of special cases.
The first (and trivial) possibility  is that the bulk of synchrotron
radiation is above the Klein-Nishina cutoff frequency. Then $y =
0$ by definition, and $e_{sy}(0)/e_{B}(0) = \Lambda \beta_{d}/2$. In the
second case, the synchrotron radiation is scattered off electrons
in the Thomson regime, while the scattering of the comptonized photons is suppressed by the Klein-Nishina effect. Then
$e_{sy}(0)/e_{B}(0)=\bar{y}_0$. Since the geometrical factor is a
few, a further approximation is not possible and Eq.~(\ref{Wsy}) has
to be solved numerically. The resulting values are $\bar{y}_0 =
0.28, 0.46, 0.65$ for $\Lambda = 3,6,10$ and $\beta_{d} =1/3$. 

Finally, it is possible that $k$ times comptonized radiation is still below the Klein-Nishina cutoff, while the next comptonization step sends photons above the cutoff. Then one has to substitute
$\bar{y}_0=(e_{sy}(0)/e_{B}(0))^{k+1}$. The actual parameters in
GRB shocks place them between the first and the second cases
above. However, relativistic shocks are able to self-tune their
parameters, as discussed in Sect.~\ref{sec:tuning}, so that they tend
towards the second case, with comptonization just approaching the
Klein-Nishina regime. The self-tuning effect has limited
capabilities and not every shock can pull its parameters to this
optimum, but for the GRB shocks this is within reach.

{Combination of large Compton $y$ parameter and fast cooling at the same time means that the IC radiation, produced in the Klein-Nishina regime or close to it, will be strongly absorbed within the shock.}
The optical depth for photon-photon collisions is
\begin{equation}
\label{tau_c} \tau_{c} = \sigma_{\gamma\gamma} n_{ph} \frac{R}{\Gamma} =
\sigma_{\gamma\gamma} \frac{R}{\Gamma} \frac{e_{B}(0)}{E_{p}},
\end{equation}
where $n_{ph}$ is the target photon number density and $\sigma_{\gamma\gamma}$ the effective two-photon pair production cross section. Here we have assumed that all the magnetic energy is eventually radiated away and we have used the magnetic energy density, $e_{B} $,  to estimate the radiation energy density. We  have ignored the logarithmic increase of the radiation energy density close to the shock front. The conservative estimate (\ref{tau_c}) assumes that the most efficiently absorbing target photons are those at the peak of synchrotron SED. 
This holds only for sufficiently hard low-energy SED asymptotes, so that in some cases the actual optical depth may be larger.

The estimated value of the optical depth for GRBs' external shocks is
\begin{equation}
\tau_{c} \simeq 
50\, {\sigma_{\gamma\gamma}}_{,-25}\, 
\epsilon_{B} R_{16} \Gamma_2^2 n_3\, \frac{m_{e} c^2}{{E_{p}}_{,lab}},
\end{equation}
where ${E_{p}}_{,lab} = \Gamma E_{p}$ is the lab-frame synchrotron peak energy. 
For internal shocks, the optical depth is 
\begin{equation}
\tau_{c} \simeq 
100\, {\sigma_{\gamma\gamma}}_{,-25}\, 
\frac{{L_{iso}}_{,51}}{\Gamma_3^2 R_{13}}\, \frac{m_{e} c^2}{{E_{p}}_{,lab}},
\end{equation}
where $\Gamma$ is the jet Lorentz factor  (the shock is assumed to be mildly relativistic). The local particle number density was expressed using the jet's equivalent isotropic power $P_{iso}$, which is related to the equivalent isotropic luminosity, $P_{iso} = L_{iso}/ \epsilon_{B}$ (assuming radiative efficiency $\simeq \epsilon_{B}$, since 
in this model the magnetic energy is the primary cause for particle heating and radiation); ${L_{iso}}$ is the equivalent isotropic luminosity. 
The estimated optical depth for absorption of IC photons is larger than unity at the early
afterglow phase and yet much larger for internal shocks.

\section{Upstream modification}
\label{sec:preacceleration}

The pairs produced through photon absorption in the upstream deposit energy and momentum,
thus decelerating the plasma flow and decreasing the difference of
Lorentz factors across the shock front. A quantitative treatment
of this process is possible under the assumption that the flow is
adiabatic, i.e., the energy and momentum of the newly born pairs are
kept at the place of their birth and neither transported 
nor re-emitted. Let us consider two planes, parallel to the shock;
one  is infinitely close to the shock front (denoted 1) and the other is
sufficiently far in the upstream (denoted 2), so that the IC radiation from
the downstream is entirely absorbed between the planes. The
upstream flow further away from the shock is undisturbed. In the
quasi-stationary case, the total energy and momentum between the
planes is conserved; this means that their fluxes, taken at the two
planes, must match each other. The corresponding equations read
\begin{equation}\label{momentumContinuity}
w_1 \beta_1^2 \Gamma_1^2 + p_1 = w_2 \beta_2^2 \Gamma_2^2 + p_2 +
S_{mom}
\end{equation}
for momentum flux and
\begin{equation}\label{energyContinuity}
w_1 \beta_1 \Gamma_1^2 = w_2 \beta_2 \Gamma_2^2 - S_{en}
\end{equation}
for energy flux. Here $w$ is the specific enthalpy, $p$ the pressure,
$\Gamma$ the bulk Lorentz factor, and $\beta$ the bulk velocity
divided by the speed of light. Equations
(\ref{momentumContinuity}) and (\ref{energyContinuity}) are nearly
the same to those involved in the derivation of the Taub adiabat, except
that their right-hand-side contains additional terms that take
into account the energy and momentum flux densities, $S_{en}$ and $S_{mom}$,
carried by the IC radiation across plane 1.

It is convenient to introduce the following parametrization:
\begin{equation}
\begin{array}{l}
S_{en} = a\, w_2 \beta_2 \Gamma_2^2 \\
S_{mom} = b\, S_{en}. \\
\end{array}
\end{equation}
Here the absorption parameter $a$ is the fraction of energy flux
at the shock front, which is injected back into the upstream, while the parameter
$b$ is the ratio of momentum
flux (multiplied by the speed of light) to the energy flux,
\begin{equation}
\int_0^{\pi/2} \cos \theta\, I(\theta) \sin \theta \, {\rm d}
\theta = b \int_0^{\pi/2} I(\theta) \sin \theta \, {\rm d}
\theta\, .
\end{equation}
For a geometrically thin shock, $b \simeq 1/\Lambda \ll 1$.

\begin{figure}
\includegraphics[width=\columnwidth]{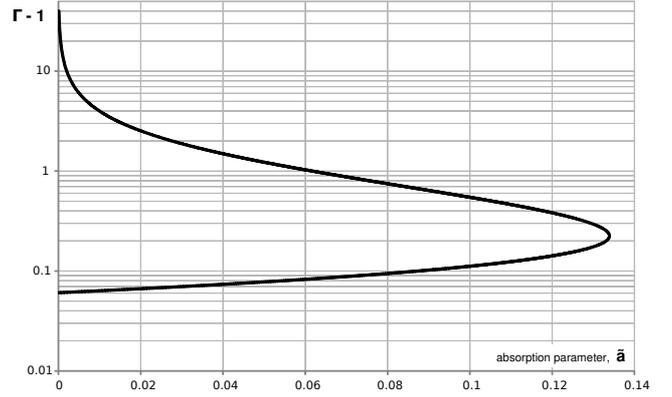}
\caption{Bulk flow Lorentz factors in the shock-front comoving frame as functions of the effective absorption parameter $\tilde{a}$. The upper branch is for the upstream and the lower branch is for the downstream. The shock Lorentz factor is infinitely large.}
\label{upstream_accel}
\end{figure}

We assume that the upstream acquires enough heat from pairs to
guarantee a relativistic equation of state ($p_2 =
w_2/4$) near the shock at plane 1, and that the shock is strong enough to make $p_1$
negligible. Then, the product $w_2 \beta_2 \Gamma_2^2$ can be
obtained from Eq.~(\ref{energyContinuity}) and substituted into
Eq.~(\ref{momentumContinuity}), resulting in:
\begin{equation} \label{velocityEquation1}
3\beta_2+\frac{1}{\beta_2} = 4(1-a)\beta_1 - 4ab \equiv 4f ,
\end{equation}
where we introduce the feedback parameter $f$, which equals
$\beta_1$ when there is no energy and momentum injection from the
downstream to the upstream. Given that $(1-\beta_1)b \ll 1$, it is
possible to eliminate $b$ by introducing an effective absorption
parameter $\tilde{a}=a(1+b)$, so that $f=(1-\tilde{a})\beta_1$.
The roots of the above equation,
\begin{equation}\label{modifiedShockVelocity}
\beta_2=\frac{2f \pm \sqrt{4f^2-3}}{3},
\end{equation}
correspond to the cases where the shock front is inside the
region between the two planes (smaller value) or outside it
(larger value, which is apparently the upstream velocity at the
shock front). The smaller is the feedback parameter, the smaller
is the upstream velocity at the shock front. There is a critical
feedback parameter $f_{cr}=\sqrt{3}/2$, such that the
discontinuity disappears; this happens already at a fairly small
absorption parameter $\tilde{a}_{cr} \simeq 0.13$. If the
efficiency of inverse Compton absorption is smaller than
$\tilde{a}_{cr}$, then the upstream Lorentz factor at the
discontinuity remains (at most) mildly relativistic down to
$\tilde{a} \simeq 0.05$ and grows as $\Gamma_2 =
1/\sqrt{4\tilde{a}}$ in the limit $\beta_1=1, \tilde{a}
\rightarrow 0$. The influence of pair loading on the
upstream flow can be considered negligible only if $\tilde{a} \ll
(2\Gamma)^{-2}$; for external shocks this is unreasonably small
value. The bulk Lorentz factors at the shock as functions of the absorption parameter 
are presented in Fig.~(\ref{upstream_accel}).

Under realistic conditions, the pre-shock region cannot be
fully adiabatic: the secondary pairs create turbulent magnetic field in
the upstream that is growing towards the shock front, and the injected electrons eventually lose some energy either through synchrotron radiation, or through inverse Compton radiation in the radiation field of the shock. 
Radiative energy losses mean that the upstream is losing its inertia
and it becomes more susceptible to deceleration, so that the upstream
Lorentz factor at the discontinuity is smaller than in
the adiabatic model.

An approximate solution for the upstream Lorentz factor (and then
for energy density and electron distribution function) can be
obtained if the equation of state for the upstream plasma is relativistic. This is satisfied for ultra-relativistic external shocks, where the upstream inevitably attains such an equation of state as it accelerates. This approximation may be less accurate for internal shocks that are only mildly relativistic.
We assume that the equation of state for the upstream plasma is relativistic (this includes turbulent magnetic field), the momentum absorption parameter $b \ll (2/\sqrt{3} - 1)$, and we treat the absorption
parameters $a = S_{en}/ w \beta \Gamma^2$ and $b=S_{mom}/S_{en}$ as functions of the distance into the upstream (for example, the absorption parameters are the fraction of energy absorbed in the upstream from some distance up to infinity). 
Then, placing the two planes (1 and 2 mentioned earlier) infinitely
close to each other, we obtain: 
\begin{equation} \label{velocityEquation2}
{\rm d} \left( 3\beta+\frac{1}{\beta} \right) = -\left(
3\beta+\frac{1}{\beta} \right) {\rm d} \tilde{a}.
\end{equation}
The differentiated value in the left-hand-side has a minimum at
$\beta=1/\sqrt{3}$ (the relativistic speed of sound), so that
$\tilde{a}$ at this point should also have an extremum.
Solving Eq.~(\ref{velocityEquation2}) for the upstream velocity in the shock-front frame, $\beta$, we obtain
\begin{equation} \label{UpstrSolution}
\left( 3\beta+\frac{1}{\beta} \right) = 4 \exp \left( -\tilde{a} \right).
\end{equation}

Returning now to the solution of the original equations we note that the energy flux continuity (Eq.~\ref{energyContinuity}) and the proton number flux continuity yield 
\begin{equation} \label{AlternContinuity}
\frac{4}{3} e_{t,u} \beta \Gamma_{u}^2 (1 - a) = \rho_0 c^2 \Gamma^2\, ,
\qquad
\beta \Gamma_{u} n = n_0 \Gamma\, ,
\end{equation}
where $\Gamma_{u}$ is the upstream Lorentz factor relative to the shock front, $e_{t,u}$ the comoving-frame energy density, $n$ the comoving number density of protons, $\rho_0$ the density of ambient medium, and $n_0$ its number density.
In the region, where the upstream is highly relativistic both in the laboratory and in the shock frame, i.e., in the limit $\Gamma \gg \Gamma_{u} \gg 1$, equations (\ref{UpstrSolution}) and (\ref{AlternContinuity}) become 
\begin{equation} \label{UpstrLoading}
\Gamma_{u} = \frac{1}{2 \tilde{a}^{1/2}}, \qquad
e_{t,u} = 3 \Gamma^2 \rho_0 c^2 \tilde{a}\, , \qquad
n=2\Gamma n_0\, \tilde{a}^{1/2}\, .
\end{equation}
{Although these expressions do not provide explicit dependence on the distance to the shock front, they are sufficient to calculate the energy distribution of electrons injected into the downstream at the shock front. }

\section{Particle injection from the upstream}
\label{sec:UpstrAccel}

The energy fed into the upstream, 
$e_{t,u} / n =(3/2) \Gamma m_p c^2\, \tilde{a}^{1/2} \propto \Gamma_{u}^{-1}$ per proton, is due to injected electrons. Their comoving-frame energy is of the order of $\Gamma_{u} E_{ic}/2$ or larger (for those injected earlier), so that the number (per proton) of electrons injected with Lorentz factor $\gamma_{i}$ or larger is $N(\gamma_{i}) \propto \gamma_{i}^{-2}$.
After injection, the electrons gain energy as a result of flow compression, and, in the absence of radiative losses, their comoving-frame Lorentz factors grow as $n^{1/3}$. By the time an electron reaches the shock front, its Lorentz factor becomes $\gamma_{f} \sim \Gamma_{u}^{1/3} \gamma_{i} \propto \gamma_{i}^{4/3}$. Then, if all these electrons were able to keep their energy rather than radiate it away, the distribution of injected electrons at the shock front would be 
\begin{equation}\label{UpstrInjection}
	N (\gamma_{f}) \propto \gamma_{f}^{-3/2}\, .
\end{equation}
This distribution begins at 
$\gamma_{f} \sim \Gamma_{sh}^2 \gamma_{cr} /4$ and cuts off at 
$\gamma_{f} \sim \Gamma^{4/3} \Gamma_{sh}^{2/3} \gamma_{cr} /4$, where $\Gamma_{sh}$ is the upstream Lorentz factor at the shock front. The total injected power can be estimated as $(\Gamma_{sh}^2 a)$ fraction of the shock power. For small absorption parameters, $\tilde{a} < 0.1$, this corresponds to a nearly constant fraction of 1/4.

The injection (\ref{UpstrInjection}) represents the response of the upstream to the peak of downstream's IC radiation. It has a power-law tail and electrons from this tail produce high-energy IC radiation, extending well beyond the IC peak. These IC photons are  absorbed again in the
upstream and  produce another generation of pairs. They are picked up by the relativistic (relative to the shock front) flow. When the pairs are brought back to the downstream, their energy is multiplied by a 
factor of at least $\Gamma_{sh}^2$ (up to $\Gamma^{4/3} \Gamma_{sh}^{2/3}$ for {a small number of pairs injected at the very beginning of the upstream acceleration}), and they will produce even more energetic IC photons. This sequence of events starts the lepton cycle of
converter acceleration \cite{ConvAcc}, which -- if not limited somehow -- deposits almost all the
available energy at the high-energy end of particle distribution,
eventually leading to a cascade and formation of a nearly flat and
featureless spectra\footnote{It should be noted that in AGNs, where the jet's Lorentz factor is moderate,  unrestricted converter acceleration may produce a reasonable outcome \cite{ConvAcc3}.}. 

Our model avoids this problem. 
As a result of the shock's self-tuning, the electrons, producing IC emission at its peak, radiate at the verge of Klein-Nishina regime and their Compton $y$ parameter is of order unity. Thus, the efficiency of IC radiation decreases as electrons are accelerated and gain energy, and the converter acceleration becomes inefficient when the electrons' Lorentz factor satisfies the condition $\Gamma_{sh}^2 y(\gamma) < 1$. For this suppression mechanism to work, one needs that $\Gamma_{sh}$ is not very large (implying a substantial upstream pair loading) and that the Compton $y$ parameter rapidly decreases with increasing electron Lorentz factor (implying a paucity of low-frequency photons in the shock's spectrum that  is achieved by heating in the downstream that  efficiently eliminates low-energy electrons).
In any case, the  injection into the shock front of a very energetic electron population with a distribution  given by Eq.~(\ref{UpstrInjection}) is an inevitable residue of the converter acceleration, which is always present if two-photon absorption opacity is not negligible. This population essentially produces a high energy transitional component in spectra, which we discuss in Sect.~\ref{sec:other_regions}.

\section{Magnetic field generation and decay}
\label{sec:Bevolution}

At the moment of their creation in the upstream, the secondary pairs have anisotropic distribution. The  recently injected pairs, whose momenta have not yet isotropized, can be considered as a ``beam'' in isotropic background plasma. The energy density of this beam is a small fraction $A \ll 1$ of the total energy density. 
The dominant contribution to the beam arises from  the 
high-energy tail of IC spectrum, where the photons' absorption timescale is of the order of $t_{abs} = l_{abs}/c$ (not taking into account the difference of bulk Lorentz factors across the shock front, which, as we have shown above, is unlikely to be large). The more numerous less energetic pairs have a lower opacity and hence a lower injection rate. Consequently,  their contribution to the anisotropic part of the distribution is less important.

The anisotropy of particle distribution function creates condition for the Weibel instability. For small anisotropy, its increment can be estimated as (\cite{WeibelIncrement}) 
\begin{equation} \label{increment}
	\Im \Omega \simeq A^{3/2} \omega_{p}.
\end{equation}
The increment peaks at wavenumber
\begin{equation} \label{Kpeak}
	k_{p} \simeq A^{1/2} \omega_{p}/c\, ,
\end{equation}
which sets the spacial scale of the magnetic field in the upstream. 

{As the magnetic field grows, the isotropization timescale becomes smaller, that means smaller anisotropy and hence smaller increment. The growth of the magnetic field saturates when the increment becomes of the order of inverse injection timescale, that is
\begin{equation} \label{weibel_A}
	\Im \Omega \sim t_{abs}^{-1}
	\qquad \Rightarrow \qquad 
	A \sim \left(t_{abs} \omega_{p}\right)^{-2/3}.
\end{equation}
The saturated magnetic field is in equipartition with the beam of injected pairs \cite{MagnStruct}, so that the upstream magnetization is
\begin{equation}
	\epsilon_{B,u} \sim A \sim \left(\frac{l_{s,u}}{l_{abs}} \right)^{2/3}	,
\end{equation}
where $l_{s,u} = c/\omega_{p}$ is the skin length of the upstream plasma. This estimate suggests that as a consequence of the  prolonged beam injection the upstream is 
only weakly magnetized, with $\epsilon_{B,u} \sim $~few$\times 10^{-3}$ at most. On the contrary, anisotropy forms very fast at the shock front and the magnetic field at the downstream side is likely to be close to equipartition with the total energy density. 
Still this initial weak component is important as it produces large scale modes that eventually determine the rate of the magnetic field decay.   
}

Estimating the anisotropy from Eq.~\ref{weibel_A} and substituting it into Eq.~\ref{Kpeak}, we obtain
the typical wavenumber for the magnetic field structures in the upstream 
\begin{equation} 
	k_{p} \sim \left(l_{abs} l_{s,u}^2\right)^{-1/3}.
\end{equation}
When amplified at the shock front, the long-wavelength modes have smaller increment, but they start from a finite seed amplitude, whereas the faster growing short-wavelength modes are dumped in the upstream to nearly vanishing amplitude. Thus, the amplified magnetic field  at the shock front roughly preserves its spacial scale. For large-scale magnetic field perturbations it takes longer for the Weibel instability to develop and the shock front width increases by a large factor from standard the value $\sim l_{s}$ to 
$\sim (l_{abs} l_{s}^2)^{1/3}$, but still remains many orders of magnitude shorter than any other spacial scale.
More importantly, the scale of perturbations determines how long the magnetic field survives in the downstream.  Starting from the shock front and until the magnetic field spatial scale becomes extremely large, 
the magnetic field decay is governed by phase mixing and the decay length is $l_{d} \sim (k_{p}^3 l_{s,u}^2)^{-1}$ (see, e.g., \cite{MagnEvolution}, but nonlinear corrections may increase the damping rate \cite{Lemoine}), which agrees with our earlier assumption $l_{d} \sim l_{abs}$. Numerical simulations also show that in the case of long particle injection the magnetic field decay lasts for a time approximately equal to duration of the injection (\cite{numeric}).

We do not derive the exact law of the magnetic field decay, but instead introduce
 two different models of the magnetic
field decay. Model {\bf I} describes exponential decay of the
magnetic field, while model {\bf II} is for a power-law decay:
\begin{equation}
\label{BdecayModels}
\begin{array}{lll} {\rm I.} \quad B = B_0 \exp(-r/l_{d}) &\Rightarrow& \gamma_0^2 \propto B^2\\[1ex]
\displaystyle {\rm II.} \quad B = B_0 \left( \frac{1}{1+r/l_{d}} \right)^{1/q} &\Rightarrow& \gamma_0^2 \propto B^{2+q}.
\end{array}
\end{equation}
{As shown in Sect.~\ref{sec:downstream_region}, different decay laws result in different low-frequency asymptotics of the downstream synchrotron emission.}

\section{Self-tuning}
\label{sec:tuning}

It is important to explore a self-regulating mechanism that arises naturally because of the downstream emission.  The  synchrotron SED around the peak is dominated by  electrons that  are close to the shock front.
The peak is located at the energy $E_{p}
\simeq \gamma_0^2(0) \hbar \omega_{B}$ (in the shock comoving frame).  The value of
$\gamma_0(0)$ (i.e., average Lorentz factor at the shock front)
can be obtained from Eq.~(\ref{balance}). A simple estimate
yields\footnote{In this section we stick to the thermal distribution as a representative example. Power-law-type distributions result in a similar qualitative behavior and in numerical factors in the following equations.}:
\begin{equation}
\label{balanceGamma} \frac{16}{9 \beta_{d}}\, \gamma_0^2(0) \sigma_{T} n_{e}
(1+\bar{y}_0) = \frac{1}{l_{d}} \ . 
\end{equation}
Substituting $l_{d}$ from Eqs.~(\ref{Bdecay}) and
(\ref{skin}) results in 
\begin{equation}
\begin{split}
\label{SEDpeak} E_{p} \simeq \frac{0.4\,
\epsilon_{B}^{1/2}}{\alpha_{f} \lambda (1+\bar{y}_0) M }\, m_e
c^2\, ,
\end{split}
\end{equation}
where $\alpha_{f}$ is the fine structure constant and
$M=n_{e}/n_{p}$ the pair multiplicity. The latter is a free
parameter, limited to $1 \leq M \lesssim m_{p}/m_{e}$, that takes into account 
possible pair loading. Given a bulk Lorentz
factor $\Gamma \sim 300$, typical to GRBs,  and  an initial magnetic energy fraction
$\epsilon_{B} \sim 0.1$, the product $\lambda M$ must be of the
order of $10^4$ for internal shocks and $10^6 \div 10^7$ for
external shocks to fit the observations of the GRB prompt emission and early afterglow emission, respectively.

Now, consider a shock, whose pair multiplicity is $M=1$
and $\lambda$ is sufficiently small, so that $E_{p} \gtrsim
m_{e} c^2/\gamma_0$, i.e. the comptonization of the synchrotron radiation proceeds in the Klein-Nishina regime or close to it. 
We will show elsewhere that self-tuning can be reached also if this condition is not satisfied. However, 
in this paper we assume that it holds. In particular this condition is met in GRBs, both for external and internal shocks. 
In this case, the typical
energy of the IC photons satisfies $E_{ic} \sim
\gamma_0 m_{e} c^2$, enough to allow two-photon pair creation in
collisions between synchrotron and IC photons, since
$E_{ic} E_{p} \gtrsim (m_{e} c^2)^2$ \footnote{Collisions between two IC photons can still produce pairs even if this
condition is not met, but the efficiency is virtually negligible.}.
Given the Klein-Nishina suppression, the power of the  IC emission
relative to that of synchrotron emission is $P_{ic} /P_{sy} =
\bar{y}_0/(1+\bar{y}_0)$ and each electron in the downstream
produces on average
\begin{equation}
\label{Minitial} M_{ph} \sim \frac{\bar{y}_0}{1+\bar{y}_0}\,
\frac{l_{d}}{l_{c}}
\end{equation}
high-energy (inverse Compton) photons. After escaping from the
shock front to the upstream, these photons may
collide with synchrotron photons, producing electron-positron
pairs.

{The pairs} are picked up by the plasma flow and are carried back to
the shock.
Thus, if $M_{ph}$ exceeds one half, the pairs
pile up in the shock, increasing the multiplicity $M$ and reducing both the average Lorentz factor $\gamma_0$ and the energy of seed photons $E_{p}$. This leads to even greater inflow of secondary pairs to the shock, since the IC
photons become less energetic (and more numerous, accordingly) and they  interact with target photons that are  closer to the cross-section maximum. 
The process  continues until the product $E_{ic} E_{p}$ drops below $(m_{e} c^2)^2$,  making pair production impossible for the bulk of IC photons. The pairs are also produced in the downstream, adding to the pair multiplication factor, but this is readily compensated by the self-tuning mechanism and thus has virtually no influence on the resulting spectrum.

Eventually, the shock reaches an attractor solution, where comptonization of synchrotron photons proceeds in the Thomson regime (so that $E_{ic} = \gamma^2 E_{p}$) and internal absorption is efficient only for the high-energy tail of the IC spectrum. The contribution of the less numerous high-energy IC photons to the total number of secondary pairs produced at the shock may be relatively small. But these photons are absorbed quickly, as the shock opacity for them is defined by Eq.~(\ref{tau_c}), and, because of their short life cycle, it is these photons that determine the pair multiplication increment, as well as the magnetic field build-up scale and strength.

At the attractor solution, the shock
parameters will tune to make $\gamma_0(0) \simeq \gamma_{cr}$, where the critical electron Lorentz factor satisfies 
\begin{equation} \label{SelfTuningCondition}
\gamma_{cr} E_{p} = \gamma_{cr}^3 \hbar \omega_{B} = m_{e} c^2,
\end{equation}
so that the rate of pair creation decreases.
Solving this relation, we obtain
\begin{equation}
\gamma_{cr} = \left( \frac{B_{cr}}{B(0)}
\right)^{1/3} \quad \Rightarrow \quad E_{p} \sim \left(
\frac{2\, L_{iso}}{B_{cr}^2 \Gamma^2 R^2 c}
\right)^{1/6} m_{e} c^2\, ,
\end{equation}
where $B_{cr} \simeq 4.5 \times 10^{13}$~G is the Schwinger field
strength. 
The lab-frame position of the synchrotron SED peak settles at
\begin{equation}
\label{tuning} {E_{p}}_{,lab} = \Gamma E_{p} 
\sim 400\, {\rm keV} \times
\frac{\Gamma_3^{2/3} {L_{iso}}_{,51}^{1/6}}{R_{13}^{1/3}}\, .
\end{equation}
This result is in remarkable agreement with the observed positions of SED peaks for both GRB prompt emission {and the early afterglows}. The tempting
closeness of the prompt emission peak energy to $m_{e} c^2$,
according to the proposed model of relativistic shock emission, is
a mere coincidence, having its roots in the fact that the
comoving-frame Lorentz factors of emitting electrons and the bulk
Lorentz factor are of the same order.

The downstream region is also producing efficiently IC radiation that, as long as the shock is in the self-tuning regime, peaks at 
\begin{equation}
{E_{ic}}_{,lab} \sim \Gamma^2
\frac{(m_{e}c^2)^2}{{E_{p}}_{,lab}} \sim 600\, {\rm GeV} \times
\frac{\Gamma_3^{4/3} R_{13}^{1/3}}{{L_{iso}}_{,51}^{1/6}} \ .
\end{equation}
However, there is no clear evidence for the second (high-energy) peak in the spectra of GRB prompt emission.  This implies that an efficient absorption mechanism, which is unrelated to the proposed shock model and operates at energies down to 10-100 GeV must be active, somewhere near the prompt emitting region,  for this model to fit observations. At the very early afterglow phase, the IC peak is expected to appear at somewhat higher energies (at or above 1 TeV) that are  greatly attenuated by the interaction with the cosmic infrared background, so that its presence may be hard to observe.

During the later afterglow phase, expansion of the external shock
eventually makes the optical depth for two-photon pair production
too small to maintain the self-tuning of the shock. From this moment, the pair loading decreases
and the electrons in the downstream increase their average Lorentz
factor to maintain balance between the heating and the losses. This creates a feature in the  lightcurve that is  possibly related to plateau behavior, observed in some GRB afterglows (see, e.g., \cite{plateau}). After the shock ends the self-tuning regime, the peak of IC emission shifts to higher energies and its power drops because of the Klein-Nishina suppression.

\section{The downstream emitting region}
\label{sec:downstream_region}

A  good proxy for the scale of the magnetic field build-up is
the photon absorption length $ l_{abs} = R/\Gamma\tau_{c}$.  Using  Eq.~(\ref{tau_c}):
\begin{equation} \label{BuildUpScale}
l_{abs} = \frac{E_{p}}{\sigma_{\gamma\gamma} e_{B}(0)}
= \frac{4 \sigma_{T}(1+\bar{y}_0)}{\sigma_{\gamma\gamma}} 
\frac{\gamma_0(0)E_{p}}{3 \beta_{d} m_{e} c^2}\,\, l_{c} \sim 20\, l_{c}\, .
\end{equation}
Here $e_{B}$ was taken from Eq.~(\ref{CoolingScale}), Eq.~\ref{SelfTuningCondition} was used to substitute $\gamma_0(0)E_{p}$, and the numerical value
for the two-photon pair production cross section used was $\sigma_{\gamma\gamma} = 10^{-25}$~cm$^2$ (close to the maximum). Given that the magnetic field dissipates on the same scale as it builds up, this guarantees that self-tuning automatically sets the shock in fast cooling regime.

We can turn now to estimate the spectrum of the emission 
produced in the downstream. For  an isotropic electron distribution the synchrotron 
emission coefficient is
\begin{equation}
\label{SySpectralEmissivity} j_{\omega}(r) = \frac{\sigma_{T}
n_{e}}{24\pi^2}\, \frac{m_{e}c^2}{e}\, B\, F_{tot} \left(
\frac{2\omega}{3 \gamma_0^2 \omega_{B}} \right) \ ,
\end{equation}
where $\omega_{B} = eB/(m_{e}c)$ is the electron gyrofrequency and
$F_{tot}(x)$ the dimensionless distribution-averaged emissivity
function
\begin{equation}
\label{Fthermal} F_{tot}(x) = \frac{1}{n_{e}} \int_0^{\infty}
\bar{F} \left( \frac{\gamma_0^2}{\gamma^2} x \right)
f_{e}(\gamma,\gamma_0)\, {\rm d} \gamma\, \ ,
\end{equation}
calculated using the distributions given in  (\ref{distribution}) or (\ref{distribution_PL}). 
Here we assume $\gamma_0 \gg 1$ in order to set 
 the limit of the integral to 0 (instead of 1).  The function  
$\bar{F}
(x)$ is the emissivity function averaged over pitch angles
\begin{equation}
\label{AngAvgF} \bar{F}(x) = \int_0^{\pi/2} \sin^2(\phi) F
\left( \frac{x}{\sin(\phi)} \right) {\rm d} \phi\, \ .
\end{equation}

\begin{figure}
\includegraphics[width=\columnwidth]{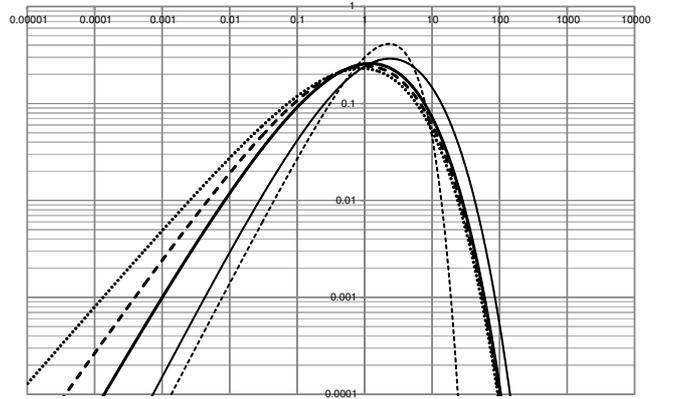}
\caption{The spectral energy distributions (in arbitrary units) of the synchrotron
radiation from the downstream region for a thermal distribution 
and for the different magnetic field
decay models (see Eq.~\ref{BdecayModels} in the text). Thick
lines: model {\bf I} (solid), model {\bf II} with $q=1$ (dashed),
and model {\bf II} with $q=2$ (dotted). For comparison, we plot
the spectral energy distributions produced in a uniform magnetic
field by a thermal distribution (thin solid line) and by
monoenergetic electrons with $\gamma=1.4 \gamma_0$ (thin dashed
line). All spectra are normalized to have the same total power. 
The frequency (horizontal axis) is in units $\gamma_0(0)^2 \omega_{B}$.}
\label{DownstreamSpectra}
\end{figure}

Finally, we use the well-known expression \cite{SySpectrum} for a single electron
moving perpendicular to the magnetic field lines
\begin{equation}
\label{F} F(x) = \frac{9\sqrt{3}}{8\pi} x \int_x^{\infty} K_{5/3}
(\xi) \, {\rm d} \xi \, ,
\end{equation}
where $K_{5/3} (\xi)$ is the modified Bessel function of the
second kind. With our choice of numerical factor in Eq.~(\ref{F}),
the normalization is
\begin{equation}
	\begin{array}{l}
\displaystyle \int_0^{\infty} F(x) {\rm d} x = 1, \qquad 
\int_0^{\infty} \bar{F}(x) {\rm d} x = \frac{2}{3}, \\
\displaystyle \int_0^{\infty} F_{tot}(x) {\rm d} x = \left\{
%{\setstretch{2.2}
	\begin{array}{ll}
		\displaystyle \frac{8}{9}, & \mbox{thermal distribution}\\
		\displaystyle \frac{8(p-2)}{9(p-3)}, & \mbox{power-law distribution}\, .\\
	\end{array}
%}
	\right.
	\end{array}
\end{equation}
Fig.~\ref{DownstreamSpectra} depicts (with thin dashed line) the function $j_{\omega}$ for the  thermal electron distribution. It's
low-frequency asymptotic behavior, $F_{tot} \propto x^{1/3}$, is
similar to that of functions $F$ and $\bar{F}$. At high
frequencies the thermal-averaged emission coefficient decays
rather slowly, $F_{tot} \propto x^{5/6} \exp(-3x^{1/3}/2^{2/3})$,
as compared to $\bar{F} \propto \exp(-x)$ and $F \propto \sqrt{x}
\exp(-x)$. The function $j_{\omega}$ for the power-law distribution differs from its thermal counterpart in the high-energy tail, which in this case is also a power-law, $F_{tot} \propto x^{(1-p)/2}$.

The spectrum of the downstream synchrotron radiation can be
obtained by integrating the emission coefficient along the shock
normal\footnote{Strictly speaking, one has to perform yet another integration over the local amplitude distribution of the turbulent magnetic field. However, this distribution is unknown and, for reasonable distributions, the integration affects only the distant part of  the high-energy tail of the spectrum, which is of limited importance in any case. So, we avoid this unnecessary complication.}:
\begin{equation}
\label{total_spectrum} I_{\omega} = \Lambda \int_0^{\infty}
j_{\omega}\left(B(r),\gamma_0(r)\right) \,{\rm d}r \, .
\end{equation}

For a given  $B(r)$, we need to obtain $\gamma_0(r)$ before carrying out this integration. 
In general this cannot be done analytically and it must be solved numerically.
However, if  $\bar{y} \gg 1$ (this
becomes increasingly accurate with a growing distance from the shock)
we can advance further.  Using
$e_{B} \bar{y}  \approx e_{r}(0)$ (that is accurate to
a logarithmic factor, provided inverse Compton scattering is in
the Thomson regime) we obtain a simple relation
\begin{equation}
\gamma_0^2 \propto \frac{\partial B^2}{\partial r}\, .
\end{equation}
Using this relation, it is possible to integrate
Eq.~(\ref{total_spectrum}) numerically for any given $B(r)$.
The inaccuracy in this approximate procedure results in  underestimating the high-energy tail of the spectrum, and, consequently, this shifts the spectral maximum to a somewhat lower frequency. 

\begin{figure}
\includegraphics[width=\columnwidth]{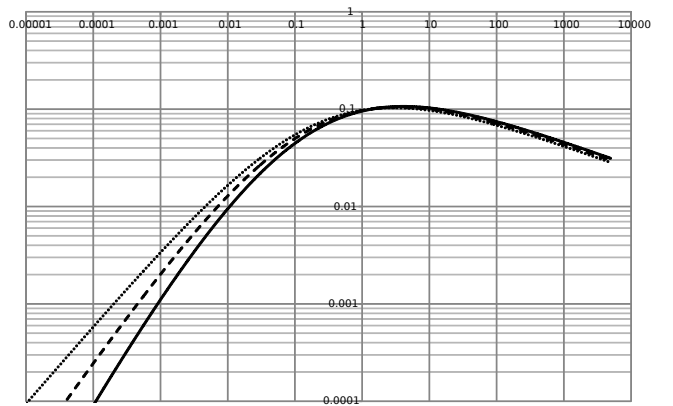}
\caption{The spectral energy distributions of the downstream region synchrotron
radiation  for a power-law distribution with $p=3.5$ 
and for different magnetic field
decay models (see Eq.~\ref{BdecayModels} in the text). Model {\bf I} -- solid line, 
model {\bf II} with $q=1$ -- dashed line, and model {\bf II} with $q=2$ -- dotted line. 
Same units and normalization as in Fig.~(\ref{DownstreamSpectra}) are used.}
\label{DownstreamSpectra_PL}
\end{figure}

The high-frequency asymptote of the integral spectrum is just the same as in the local
emission coefficient, $F_{tot}$, because it is dominated by the most energetic electron population from the vicinity of the shock front. On the contrary, the resulting
spectral index of the low-energy tail depends on the magnetic
field decay law because the contribution of less energetic electrons from distant parts of the downstream may be important. To illustrate this, we {calculate the} synchrotron spectra for {different magnetic field decay laws.} {The results} are shown in
Fig.~\ref{DownstreamSpectra} (for a thermal distribution) and Fig.~\ref{DownstreamSpectra_PL} (for a distribution with a power-law tail with an index $p=3.5$).

To reveal the low-frequency asymptote analytically, one may take
a delta function instead of $F_{tot}$ and then make use of the simple
proportionality $\omega \propto \gamma_0^2 B$ and ${\rm d}I
\propto \gamma_0^2 B^2 {\rm d}r$. In the case where $\omega$,
$\gamma_0$ and $B$, raised to appropriate powers, are all
proportional to each other, we have
\begin{equation}
\label{low-energy} \omega I_{\omega} = \frac{{\rm d}I}{{\rm d}
(\ln\omega)} \propto \frac{{\rm d}I}{{\rm d}(\ln B)} \propto
\gamma_0^2 B^3 \left( \frac{\partial B}{\partial r} \right)^{-1}
\propto B^4\, ,
\end{equation}
where $B$ is to be substituted using Eq.~\ref{BdecayModels} and the relation  $\omega \propto \gamma_0^2 B$. 
A power-law magnetic field decay (model {\bf II}) produces
low-frequency asymptote $\displaystyle \omega I_{\omega} \propto
\omega^{\frac{4}{3+q}}$, which is softer than the spectrum of an
individual electron and approaches it only in the limit of $1/q
\rightarrow \infty$ {(the power-law decay becomes exponential in this limit)}. An exponential decay (model {\bf I}) produces in
this approximation the hardest possible low-frequency spectrum,
$\omega I_{\omega} \propto \omega^{4/3}$, though a more accurate
analysis reveals the asymptote $\omega I_{\omega} \propto
\ln(\omega_{p}/\omega) \omega^{4/3}$, still slightly softer than
for an individual electron.

\section{The transitional radiation and the upstream emitting region}
\label{sec:other_regions}

From the point of view of an individual
electron, the upstream evolution looks as follows. When an energetic
electron is produced, via pair creation far in the upstream, it finds itself in a rather weak magnetic field.
Because of small magnetic field and strong Klein-Nishina suppression of IC radiation, it does not cool within the upstream dynamical timescale, but 
instead it gains energy due to adiabatic compression of the decelerating upstream flow. As time passes,
the electron's comoving-frame Lorentz factor grows and so does the magnetic field
strength around it. At the same time, the Lorentz factor relative to the shock's radiation field decreases, reducing the Klein-Nishina effect and increasing the  energy losses due to IC. For the most energetic electrons, the cooling timescale may eventually become comparable to the shock dynamical timescale. From this moment onwards,  the evolution of the electrons' Lorentz factors resembles the standard ``injection followed by fast cooling'' scheme, where a power-law like spectrum is expected. This upstream emission component is due to a small number of secondary electrons injected at the time when the upstream fluid was not accelerated to relativistic velocity. The spectrum and the overall amplitude of this component are sensitive to details of the shock structure, which are beyond the scope of this paper. A unique feature of the upstream emission component is that it is produced in those regions of the  upstream, which are close to the shock front and in which  the bulk lab-frame Lorentz factor is large; 
it is just a few times smaller than that of the downstream. This  emission that  is roughly isotropic in the comoving frame, is beamed into a cone with a larger opening angle in the lab frame than the downstream emission.  Thus this upstream emission component can be observed at  
larger angles and with a different temporal behavior compared with the downstream component originating from the same distance to the central engine.

The synchrotron cooling length for an electron in the upstream -- given its low magnetization -- is several orders of magnitude larger that the downstream cooling length, and the latter is only an order of magnitude shorter than the scale of magnetic field build-up (see Eq.~\ref{BuildUpScale}). Under these circumstances, the synchrotron radiation is negligible for the bulk of secondary pairs in the upstream. However, those produced with Lorentz factors larger than 
$\gamma_{c} \sim \gamma_{cr} \epsilon_{B} l_{c}/(\epsilon_{B,u} l_{abs})$ will be in the fast cooling regime, provided the shock Lorentz factor is large enough to reach this limit at least early in the course of upstream acceleration. The inverse Compton radiative losses are marginally efficient for the bulk of secondary pairs. Since IC cooling proceeds in the Klein-Nishina regime, its net effect depends on the shape of synchrotron SED below the peak frequency. If this part of the spectrum is harder than $\nu F_{\nu} \propto \nu$, then IC cooling timescale increases with increasing electron Lorentz factor and the radiative losses in the upstream are not important for all electrons with $\gamma \ll \gamma_{c}$.  In this case, the injected distribution (\ref{UpstrInjection}) reaches the downstream largely intact. 
At the moment of shock crossing, the distribution of secondary pairs extends well beyond the equilibrium Lorentz factor $\gamma_0(0)$ and when they enter the downstream they quickly cool producing the transitional emission component, which was mentioned earlier.
For realistic cases this distribution has a cutoff, which is more likely due to 
finite shock Lorentz factor, than due to cooling in the upstream. 
However, if the low-frequency asymptote of the synchrotron SED is softer than $\nu F_{\nu} \propto \nu$, then electrons with larger Lorentz factors have smaller IC cooling timescale. This redefines the cooling cutoff $\gamma_{c}$, decreasing it. In extreme case (very soft low-frequency asymptote) it becomes comparable to $\gamma_0(0)$ and the transitional component effectively disappears.

\begin{figure}
\includegraphics[width=\columnwidth]{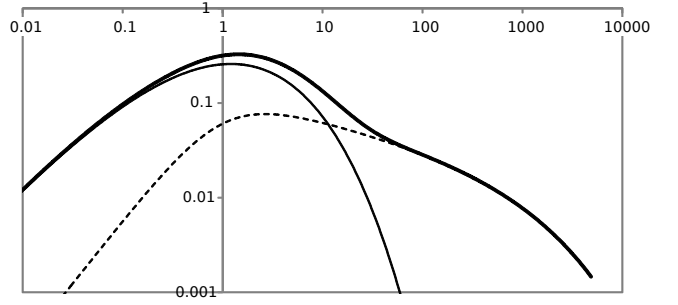}
\caption{Composite synchrotron SED (thick solid line), resulting from superposition of emission from thermal downstream distribution (thin solid line, same as shown by thick solid line in 
Fig.~(\ref{DownstreamSpectra})) 
and from fast-cooling injection at the shock front (schematically shown by thin dashed line). 
The cutoff Lorentz factor is $\gamma_{c} = 30 \gamma_0(0)$ and normalization of the fast-cooling component corresponds to average Compton parameter $\bar{y} =0.5$.}
\label{CompositeSED}
\end{figure}

Whatever is the reason for the distribution (\ref{UpstrInjection}) to cut off, it is nearly a power-law between $\Gamma_{sh}^2 \gamma_{cr} /4$ and $\gamma_{c}$ (effectively, 
$\gamma_{c} \sim \Gamma^{4/3} \Gamma_{sh}^{2/3} \gamma_{cr} /4$ 
if cooling is not important in the upstream). In the downstream, the injected electrons find themselves in the fast cooling regime if their Lorentz factor is larger than $\gamma_{cr}$ and in a ``fast heating'' regime otherwise. The fast-cooling part of the injected electrons emits synchrotron radiation with the spectrum \cite{MultiZone} 
\begin{equation}
	\nu F_{\nu} \propto \frac{\gamma_{f} N(\gamma_{f})}{1+y(\gamma_{f})}  \propto \nu^{-1/4}\, ,
\end{equation}
where the synchrotron frequency is $\nu \propto \gamma_{f}^2$ and the Compton $y$ parameter is small for electrons with $\gamma_{f} \gg \gamma_{cr}$. This slowly declining power-law spectrum extends from 
$\sim E_{p}$ to $\sim (\gamma_{c}/\gamma_{cr})^2 E_{p}$ and adds a high-energy 
tail to the shock's synchrotron SED even in the case where it is absent in the spectrum of the downstream emission. An example of such a composite spectrum is presented in Fig.~(\ref{CompositeSED}).

\section{Discussion}
\label{sec:discussion}

We have presented a general self-consistent  model for relativistic shocks with large compactness.  
Its natural applications are GRBs and AGNs, where one finds relativistic outflows and the compactness is high. The model allows us to calculate  both the shock structure and the spectrum of its radiation. 
The model involves a relatively complex network of various processes, so that to explore it analytically we have to make some simplifying approximations and even introduce  a phenomenological description at some points. Our analysis yields several important results. We list them here and briefly discuss how these results are related to GRB observations. 
 
We have shown that the standard picture of collisionless shocks should be modified due to long range coupling between the upstream and the downstream. This coupling arises due to absorption in the upstream of IC photons produced at the downstream; it regulates and influences the structure of the shock.  The shock has an intrinsic process  of self tuning. As more and more pairs are produced in the upstream their multiplicity in the downstream increases reducing the typical synchrotron and IC energies until pair production in the upstream becomes marginal.  Due to this
 self-tuning,  the shock's  parameters  evolve toward an attractor solution  in which the peak of the synchrotron SED is at such an energy that for an average radiating electron comptonization proceeds close to the Klein-Nishina regime.

Due to the deposition of momentum by the IC photons in the upstream this region accelerates. The major change in the Lorentz factor takes place
in the upstream as it is gradually loaded with secondary pairs and  the bulk Lorentz factor jump at the shock front is at most mildly
relativistic even if the shock itself is ultrarelativistic (this makes the difference between GRB internal and external shock far less dramatic). At the same time, the downstream velocity appears to be considerably larger than the value $c/3$ found in unmodified shocks.
The relatively small bulk Lorentz factor jump at the shock front combined with the Klein-Nishina suppression of the IC radiation at the high-energy tail of the electron distribution greatly diminish the efficiency of the converter acceleration discussed earlier and
the shock avoids the unacceptable SEDs produced in this mechanism. 
The diffusive shock acceleration is also suppressed, because of the  large downstream velocity, and it probably doesn't play any important role.

One of the most pronounced features of this model is that IC emission is approximately as efficient as synchrotron. 
This is an integral part of the model as these IC photons are those responsible for the self-tuning. 
There is no way to get rid of this IC peak within the model's framework. 
This emission  is
partially obscured from view by two-photon absorption within the shock itself, but the low-energy portion of the IC spectrum emerges unattenuated. 

Unlike Blazars, GRBs don't have a prominent GeV second peak. With typical parameters this second peak is expected at around 100 GeV. While the limits on the prompt GRB spectra don't extend all the way to 
100 GeV, it is clear that the lower energy ($\sim 1 \div 10$ GeV) tail of this component is not observed. 
The only way to resolve this  discrepancy is to postulate the absorption of these high-energy photons close to the source. There are reasons for such an absorption in the high-compactness zone close to the central engine.  
The appearance of this second peak is not problematic for the external shock afterglow emission as in this case it is expected to be in the TeV range that is absorbed by the intergalactic IR background.

At the prolonged afterglow phase, the conditions change. Eventually the shocks become more and more transparent
to IC radiation and at some point pair multiplication stops
(gradually, rather than abruptly), the equilibrium electron
Lorentz factor rises and so does the radiative efficiency in the synchrotron range. 
This compensates (or even overcompensates) for the effects caused by shock expansion and deceleration, 
thus producing a kind of plateau in the afterglow lightcurve. This predicted behavior may be related to the plateau feature observed in GRB afterglows. 
Despite the failure of self-tuning, the shock
evolution after this point is still governed by the same set of physical processes and can be described with the help of our model. This analysis will be presented elsewhere.

The present model includes three distinct emitting regions with different spectra and evolution. Most of the emitted power comes from an extended region in the downstream with a declining magnetic field. Another  emitting region is a thin layer next to the shock front, where energetic pairs advected  from the upstream cool rapidly. This transitional region is responsible for the high-energy power-law tail in the observed spectra; it arises due to the  converter acceleration. Finally, the most energetic electrons in the upstream cool radiatively. 
The most luminous part of the upstream region has a Lorentz factor lower than that of the downstream and hence it has a broader beaming pattern and will have a different temporal pattern.

Turning to observations we note that there are indications that the observed spectra of GRB prompt emission can be decomposed into three components \cite{3components}, one of them looks like a thermal component, possibly a photospheric emission and the other two are presumably from optically thin regions, e.g. from internal shocks.  If so, they may be related to the downstream and the transitional components in our model. Also, some afterglows show different behavior of optical and X-rays \cite{ChromBreaks}, which may be a signature of different emission components as well.

Finally we note that while the model has some promising features for the prompt spectra, such as the peak energy of the photons and the appearance of several different emission components, it does not resolve the so called ``synchrotron line of death" problem. The low energy spectra are about as good in producing hard low-frequency asymptotes as it theoretically could be: under the right choice of parameters the low-frequency asymptote of shock's SED appears to be almost as hard as that of an individual synchrotron-emitting particle. However, some of the observed prompt GRB spectra violate this limit \cite{hardsynch}. So there must be something more about the prompt phase, that is not captured by our model.

To conclude we note that the model proposed here offers a novel point of view on the physics of particle acceleration and magnetic field build up and decay in collisionless relativistic shocks. This model, which has never been explored before, ties together all the essential component of shock dynamics. The complexity of this model doesn't allow us to derive all its  details here. Hence we explore only 
a particular implementation. The validity range of this implementation is rather broad, 
and in particular it is applicable to both prompt and afterglow GRB shocks.  
Still it does not cover all the interesting parameter space. In the future, other implementations of the same model framework will be explored, extending  it to different regions in the  parameter space with possible application to other astrophysical phenomena.

\section*{Acknowledgements}

We thank Jonathan Katz for helpful comments. This research was supported in part by the Ministry of Education
and Science of the Russian Federation under Contract No.14.Z50.31.0007, a RFBR grant 14-12-00766a, a CNSF-ISF grant  394/13, an ISA grant 3-10417, by the ISF-CHE I-Core center of excellence grant 1829/12 and by a JTF grant.

%%%%%%%%%%%%%%%%%%%%%%%%%%%%%%%%%%%%%%%%%%%%%%%%%%



\bibliographystyle{mnras}
\bibliography{references}

%%%%%%%%%%%%%%%%%%%%%%%%%%%%%%%%%%%%%%%%%%%%%%%%%%

% Don't change these lines
\bsp	% typesetting comment
\label{lastpage}
\end{document}